\documentclass[prd,twocolumn,aps,superscriptaddress,nofootinbib,showpacs,preprintnumbers]{revtex4-1}
\usepackage{amsfonts,amssymb,amsmath,amssymb}
\usepackage{color}

\usepackage{bm}
\usepackage{times}
\usepackage{hyperref}
\hypersetup{
  colorlinks=true,
  citecolor=blue,
  linkcolor=blue,
  urlcolor=blue}

\usepackage{epsfig}
\usepackage{dcolumn}
\usepackage{float}

\usepackage[usenames,dvipsnames]{xcolor}
 
\definecolor{X575}{rgb}{0.05, 0.7, 0.05}

\usepackage{graphics,graphicx}

\newcommand{\ba}{\begin{array}}
\newcommand{\ea}{\end{array}}
\newcommand{\beq}{\begin{equation}}
\newcommand{\eeq}{\end{equation}}

\def\bt{\begin{table}}
\def\et{\end{table}}
\def\bc{\begin{center}}
\def\ec{\end{center}}
\def\bi{\begin{itemize}}
\def\ei{\end{itemize}}
\def\bea{\begin{eqnarray}}
\def\eea{\end{eqnarray}}
\def\beas{\begin{eqnarray*}}
\def\eeas{\end{eqnarray*}}

\def \MET{\rm E{\!\!\!/}_T}

\begin{document} 

\renewcommand{\arraystretch}{3}

\title{Unraveling the CP phase of top-Higgs coupling in associated production at the LHC}

\author{Saurabh D. Rindani}
\email[Email Address: ]{saurabh@prl.res.in}
\affiliation{
Physical Research Laboratory, Navrangpura,
Ahmedabad 380 009, India}

\author{Pankaj Sharma}
\email[Email Address: ]{pankaj.sharma@adelaide.edu.au}
\affiliation{Center of Excellence in Particle Physics at Tera Scale,
  The University of Adelaide,
  5005 Adelaide, South Australia}

\author{Ambresh Shivaji}
\email[Email Address: ]{ambresh.shivaji@pv.infn.it}
\affiliation{INFN, Sezione di Pavia,
  Via A. Bassi 6, 27100 Pavia, Italy}
  
\begin{abstract}
We study the sensitivity of top polarization observables to the CP phase $\zeta_t$ in the top Yukawa coupling in the process  $pp\to thj$ at the 14 TeV high-luminosity run of the Large Hadron Collider (HL-LHC). We calculate the top polarization in this process as well as an azimuthal asymmetry of the charged lepton arising from the decay of the top in the lab frame. We find that the dependence of this lab-frame azimuthal asymmetry on the phase $\zeta_t$ closely resembles the dependence of the top polarization on $\zeta_t$. As compared to the cross section, which is sensitive to $\zeta_t$ for larger values, the lepton azimuthal asymmetry can provide a sensitive measurement of $\zeta_t$ for smaller values.   
\end{abstract}

\preprint{ADP-16-19/T974}

\maketitle

\section{Introduction}

Particle physics has entered a new era with the discovery at the Large Hadron Collider (LHC) of a spin-0 particle of mass around 125 GeV in its first run \cite{Chatrchyan:2012xdj,Aad:2012tfa}. The couplings of this particle, presumed to be a Higgs boson, to standard model (SM) fermions and electroweak (EW) gauge bosons have been determined through the measurement of its production and decay properties, albeit with large uncertainties. Thus the current LHC data still permits a lot of leeway  for the
existence of new physics. Currently the Higgs boson couplings to the EW gauge bosons $W,Z$ point to a spin-0 particle with a purely pseudoscalar boson being ruled out at 95 \% CL \cite{Aad:2013xqa}. However a CP mixture with both scalar and pseudoscalar components is still allowed. Thus it would be one of the important goals of the next run of the LHC, which will be a high energy and high luminosity run, to determine the CP composition of the Higgs.

In this context, Higgs boson couplings to the third generation of fermions, particularly the top quark, are important since the corresponding Yukawa couplings are the largest. So far, the information regarding the $t{\bar t} h$ coupling is inferred from loop-induced $hgg$ and $h\gamma\gamma$ couplings, which are deduced from the Higgs boson production and decay at the LHC. However as these processes are loop induced, they may involve contributions from new physics. Thus, at the LHC, the top Yukawa coupling can be directly probed only in production associated with a Higgs boson as the decay $h\to t\bar t$ is kinematically forbidden. In the SM, there are two associated top-Higgs production processes possible: a) Higgs with a $t\bar t$ pair and b) Higgs with a single top, the former being the dominant one. 

In this letter, we study single-top production in association with a Higgs boson $h$ and a light-quark jet, which we denote as $thj$ production. This process has a low cross section in the SM, around 18 (70) fb at NLO at 8 (14) TeV \cite{Farina:2012xp,Demartin:2015uha}. However, in the presence of anomalous couplings, the cross section can be substantially enhanced \cite{Agrawal:2012ga}. The reason is that in the SM, there is a high degree of destructive interference between the diagrams containing Higgs emission from an internal $W$ line and from a top-quark line. If either the $WWh$ coupling or the $t\bar t h$ coupling is anomalous, the cancellation between the two types of diagrams does not take place, and the cross section is high. For example, a change in the sign of the $t\bar t h$ coupling results in a cross section of 235 fb, significantly higher than even the $t\bar t h$ cross section of 130 fb at 8 TeV \cite{Heinemeyer:2013tqa}. This allows the flipped sign of the top Yukawa coupling to be observed or excluded \cite{Farina:2012xp,Demartin:2015uha,Demartin:2014fia,Chang:2014rfa,Biswas:2012bd,Biswas:2013xva,Buckley:2015vsa}. The CMS collaboration at the LHC performed searches for this process for a variety of signatures, covering various Higgs decay channels, assuming the top quark to decay semileptonically \cite{Khachatryan:2015ota}, putting limits on the cross section. Thus, though the process of $th$ production at the LHC has negligible cross section in the SM, it can become observable when there are anomalous couplings present. In particular, the cross section is sensitive to the phase $\zeta_t$ of the top Yukawa coupling. This phase determines the pseudoscalar admixture to the scalar coupling, and is thus CP violating. It is found that increasing $\vert \zeta_t \vert$ reduces the $pp \to t\bar th$ cross section \cite{Ellis:2013yxa}, but enhances the $pp\to thj$ cross section \cite{Chang:2014rfa,Kobakhidze:2014gqa}. 

Mainly because of its large mass $m_t=172.99\pm 0.91$ GeV \cite{Aad:2015nba}, the top-quark sector is considered to be one of the few places where new physics could arise. The top-quark life time is very short and the top decays rapidly before any non-perturbative QCD effects can force it into a bound state. Thus, its spin information is preserved in terms of the differential distribution of its decay products. So by studying the kinematical distributions of top decay products, it is, in principle, possible to measure top polarization in
any top production process. As a pseudoscalar coupling violates parity, it flips the spin of the top quark when a Higgs boson is emitted. This fact has been utilized in many studies. Top-quark polarization thus depends on the phase $\zeta_t$ \cite{Ellis:2013yxa,Yue:2014tya}, and may be used to distinguish among various choices of phases. Ellis et al. \cite{Ellis:2013yxa} consider longitudinal as well as transverse polarizations as measured by the forward-backward asymmetry of the decay lepton with respect to the spin-quantization axis in the rest frame of the top. Yue in \cite{Yue:2014tya} has analyzed the utility of the $h \to \gamma\gamma$ channel as a probe of the CP-violating phase $\zeta_t$ in the process $pp \to thj$, taking advantage of the fact that in addition to the cross section and the top-quark polarization, also the branching ratio for the diphoton channel increases with $\vert \zeta_t\vert$.
 
In this work, we focus on $thj$ production in the presence of the CP-violating phase $\zeta_t$ of the top Yukawa coupling at the 14 TeV LHC and examine the possibility of using top polarization and other angular observables constructed from top decay products in the top rest frame as well as the laboratory (lab) frame to measure this phase. Since earlier work has largely focused on  measurement of cross sections and of top polarization through decay distributions in the top rest frame to enable the determination of the top Yukawa coupling and its phase $\zeta_t$, our main emphasis will be to show how lab-frame observables can be used to probe $\zeta_t$. 

The rest of the article is organized as follows. In the next section, we write down the effective top-Yukawa coupling and constraints on the CP violating pahse $\zeta_t$ from Higgs production and decay processes. In Sec.~\ref{cs} we describe the results of the calculation of the cross section for the process, and in Sec.~\ref{pol} we study the top polarization and its reconstruction through charged-lepton angular distributions in the rest frame as well as in the lab frame. In Sec.~\ref{sec-asym}, we discuss asymmetries in the rest frame of the top quark as well as in the lab frame to study their sensitivities to determine the CP phase.  Our conclusions are contained in Sec.~\ref{conclusions}.

\section{Effective top-Yukawa couplings}\label{coup}

In an extension of the SM, where there is at least one extra neutral Higgs boson, the mass eigenstates of the scalars will in general be mixtures of the original states. In case CP is not conserved, there can be mixing between CP-even and CP-odd scalars, giving rise to CP-violating couplings of the scalar eigenstates. We analyze the results of such a mixing in a model-independent scenario and parametrize the couplings in a general way.

Thus, assuming that a scalar $h$ is a mass eigenstate, the most general $t\bar t h$ coupling, without imposing CP invariance, may be  written as 
\begin{equation}\label{tthcoupling}
\mathcal L_{t\bar t h} =  - y_t \ \bar t \ (\cos\zeta_t + i \ \gamma_5 \ \sin\zeta_t)t 
\ h .
\end{equation}
Here $\zeta_t$ is the phase of the Yukawa coupling. $\zeta_t=0$ or $\zeta_t=\pi$ correspond to a pure scalar state while $\zeta_t=\pi/2$ to a pure pseudoscalar state. Any intermediate value $0<\zeta_t<\pi/2$, or  $\pi/2<\zeta_t<\pi$ signals CP violation. $\zeta_t=\pi/4$ denotes a maximally CP violating case. In this work, we focus on the effects of $\zeta_t$, so we will take $y_t=y_t^{\rm SM}=m_t/v$ while treating $\zeta_t$ as a free parameter.

Constraints on $y_t$ and $\zeta_t$ have been obtained from current LHC data. In Refs.~\cite{Nishiwaki:2013cma,Ellis:2013yxa,Kobakhidze:2014gqa,Boudjema:2015nda,Cheung:2013kla,Cheung:2014noa}, using the limits on $hg g$ and $h \gamma\gamma$ couplings derived from the Higgs boson production and decay respectively, the authors have obtained constraints in the plane of $(y_t,~\zeta_t)$. Constraints on these parameters are also derived taking into account the unitary violation in gauge boson ($W,Z$) scattering with the top quark~\cite{Bhattacharyya:2012tj,Choudhury:2012tk}. The most stringent constraints on the phase $\zeta_t$ comes from electron dipole-moment (EDM) measurements \cite{Brod:2013cka,Cirigliano:2016njn,Chien:2015xha}. These analyses are based on certain assumptions about Higgs couplings to other fermions and gauge bosons. However relaxing those assumptions can allow, in principle, a larger values for $y_t$ and $\zeta_t$. For example, in presence of only anomalous top Yukawa coupling, the current bound from electron EDM measurement allows values for the phase $\zeta_t$ in a narrow band around 0 and $\pi$~\cite{Brod:2013cka}. However, if we assume similar anomalous coupling for the electron as well, $\zeta_t$ can take any value between 0 and $\pi$ and is highly correlated with the phase corresponding to electron Yukawa ($\zeta_e$). The conclusion remains same for future prospect where the experimental bounds are expected to improve by a factor of 20 resulting into a tighter correlation between electron and top Yukawa phases. The EDM constraints from neutron and mercury atom are also expected to get much relaxed if light quark anomalous couplings are turned on. Note that assuming light fermion Yukawa couplings also anomalous does not affect our collider signal. 

On the collider side, with $y_t = y_t^{\rm SM}$ the global analysis allows $\zeta_t$ in the range [0, $2\pi/3$] at 95\% confidence level. Nevertheless, it is clear that the cases of $\zeta_t = \pi/2$ and $\zeta_t = \pi$ are already ruled out by the LHC Higgs data. The forecast for the future sensitivity at 14 TeV LHC with $3000$ fb$^{-1}$ integrated luminosity can push $\zeta_t$ very close to 0.003$\pi$. The expected senstivity at 240 GeV TLEP would be able to rule out values of $\zeta_t$ larger than $0.07\pi$ \cite{Brod:2013cka,Kobakhidze:2014gqa}. However these limits have been obtained using loop processes while the objective of the present work is to measure the CP violating phase from direct search. The existing limits on top Yukawa from the direct searches in $pp \to t{\bar t}h$ channel are very poor~\cite{Nishiwaki:2013cma,Mileo:2016mxg}.

In what follows, we assume that $h$ is indeed the spin-0 boson with a mass of about 125 GeV discovered at the LHC. Also for the sake of completeness, we vary $\zeta_t$ in the full range between $0$ and $\pi$. Since the $WWh$ coupling is directly constrained by the Higgs data, we stick to its SM value in our analysis.

\section{Signal and Backgrounds}\label{cs}

\begin{figure}[h!]
\begin{center}
\includegraphics[width=0.4\textwidth]{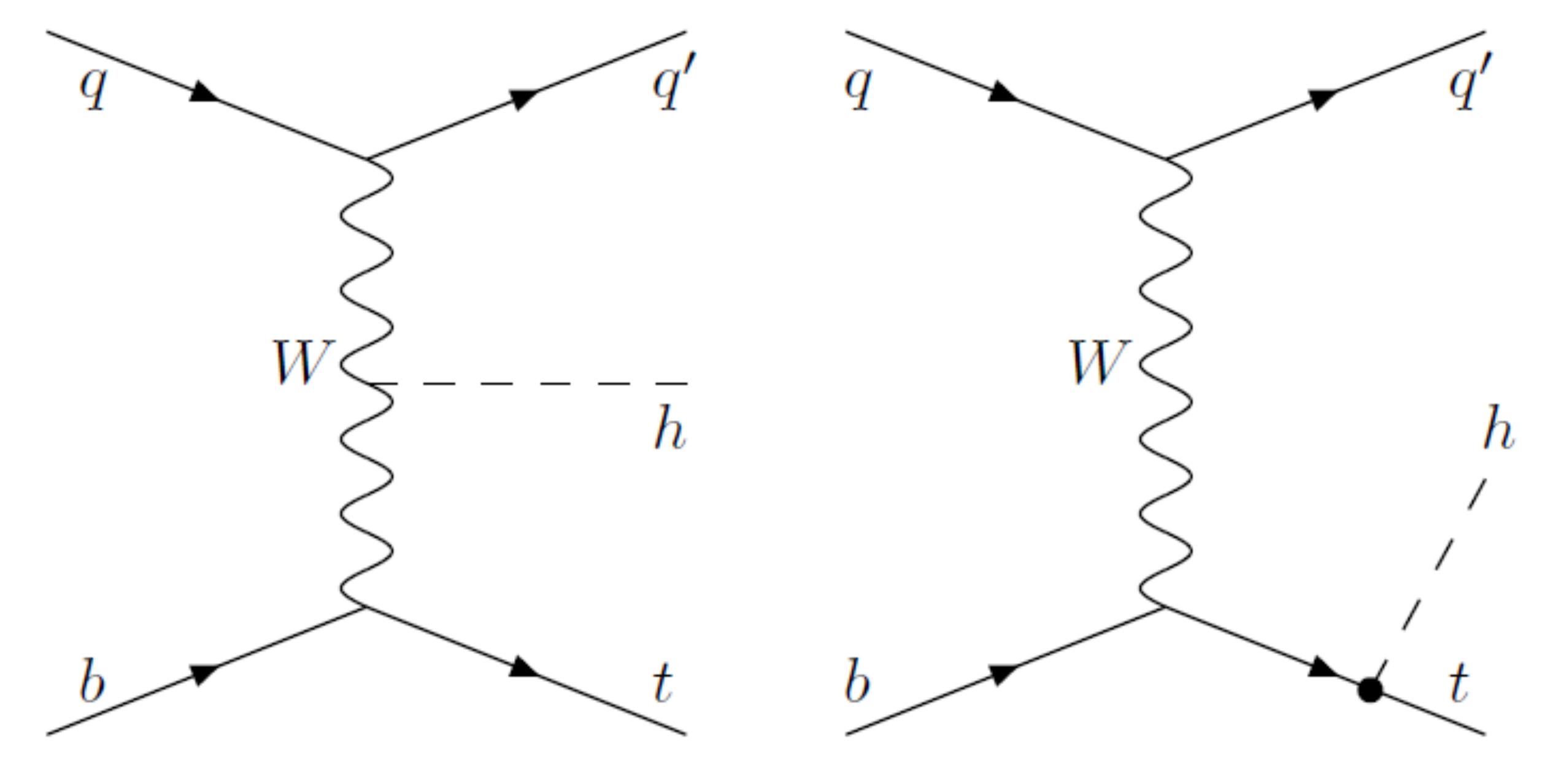} 
\caption{\label{diag}Feynman diagrams for the process $b q \to t h j$ at the LHC. The blob denotes the effective $t\bar t h$ coupling.}
\end{center}
\end{figure}

Associated production of the top quark with a Higgs and a jet at the LHC proceeds via the partonic process 
\begin{equation}\label{bqtothj}
b + q \to t + h + q',
\end{equation}
where $q$, $q'$ denote light quarks. The corresponding Feynman diagrams are shown in Fig.~\ref{diag}. As Higgs couplings to the light quarks and the $b$ quark are negligible, the corresponding diagrams are not shown. 

\begin{figure}[h!]
\begin{center}
\includegraphics[width=0.4\textwidth]{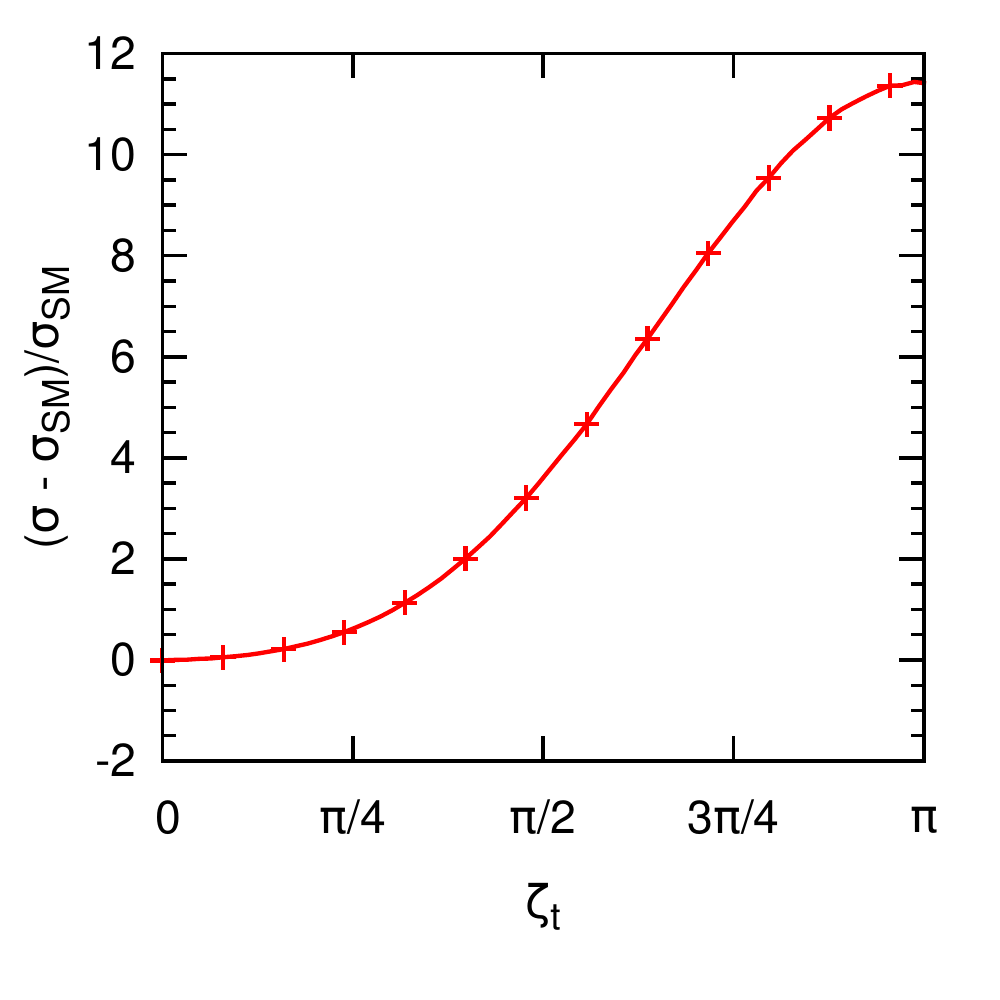} 
\caption{ The fractional deviation of the cross section from the SM value as a function of CP phase $\zeta_t$ in the $t\bar t h$ coupling for $thj$ process at LHC14.}
\label{xsec}
\end{center}
\end{figure}
We implement the effective $t\bar t h$ couplings of Eq.~\ref{tthcoupling} using {\tt FeynRules}~\cite{Alloul:2013bka} and obtained the cross section for $thj$ production for the 14 TeV LHC at the leading order using {\tt Madgraph}~\cite{Alwall:2014hca}. In Fig.~\ref{xsec}, we show the fractional deviation in the production cross section including anomalous couplings relative to the SM. We find that the cross section is fairly sensitive to the CP phase $\zeta_t$ in $t\bar t h$ couplings in the region $\zeta_t>\pi/2$ where the interference between the two diagrams becomes constructive. Below $\zeta_t<\pi/2$ the interference is still destructive though its degree decreases with $\zeta_t$, thus increasing the cross section by around 200\% at $\zeta_t=\pi/2$. On the other hand, for $\zeta_t=\pi$ the cross section can be enhanced by up to 1200\%.

Let us now consider the possible signatures of the $thj$ process at the LHC and the corresponding dominant backgrounds. The search strategy for the $thj$ signal relies on the very forward light-flavour jet which opportunely enhances the signal-to-background ratio. For the Higgs of mass of 125 GeV, the dominant decay mode is to a pair of $b$ quarks with branching fraction (BR) around 60\%. However cleanest decay mode is $h\to \gamma\gamma$ using which the Higgs was first observed at the LHC. Despite of its very small BR, {\it viz.},  $2\times 10^{-3}$, it has been shown in Refs.~\cite{Biswas:2012bd,Yue:2014tya} that the viability of the $pp\to thj(h\to \gamma\gamma)$ signal reaches a sensitivity similar to the one where the Higgs decays to a pair of $b$ quarks. The observability of the $pp\to thj$ process at the LHC in $b\bar b$ decays of the Higgs has been studied extensively in Refs.~\cite{Maltoni:2001hu,Barger:2009ky,Kobakhidze:2014gqa,Agrawal:2012ga}. As our  lab-frame asymmetry does not depend on the different modes of Higgs decay but only on the charged lepton coming from top decay, we consider both the Higgs decays in our analysis in order to enhance the statistical significance of the observables. 

For the case where $h$ decays to a $b\bar b$ pair and the top decays semi-leptonically, the signal constitutes of an isolated charged lepton $\ell^\pm$, 3 $b$ jets, 1 forward jet and missing transverse energy $\MET$. The irreducible background contribution to such a signal comes from $Wbbbj$ processes. The $Wbbbj$ processes include the contribution from
single-top processes, {\it viz.}, $tZj$ and $tb\bar b j$. The dominant background comes from top-pair production $t\bar t+j$ where one of the light jets fakes a $b$ jet. Moreover there are other QCD backgrounds resulting from light jets faking $b$ jets as in $tbjj$ and $Wbbjj$. All these backgrounds have been systematically analyzed in Ref.~\cite{Agrawal:2012ga,Maltoni:2001hu} where the authors use some standard cuts to reduce the backgrounds and improve the signal-to-background ratio.

On the other hand, when $h$ decays to a photon pair, the signal consists of an isolated charged lepton $\ell^\pm$, one $b$ jet, one forward jet, a pair of photons and missing transverse energy $\MET$. For such a signal, the irreducible background is a $tj\gamma\gamma$ continuum. As this background is non-resonant, it can be efficiently suppressed through a cut on the invariant mass of the photon pair. Other reducible contributions are from $t\bar t \gamma\gamma$ where one of the two tops decay hadronically, $b$ is mistagged as a light jet and two of the light jets do not fall inside the detector, and from $Wjj\gamma\gamma$ where one of the light jets is mistagged as a $b$ jet~\cite{Biswas:2012bd,Yue:2014tya}.

In the following, we present various angular distributions of the charged lepton coming from top decay both in the top rest frame and in the lab frame. We work at the parton level throughout, and in presenting all distributions, we apply the following standard cuts: 
\begin{eqnarray}
 p_T^{b,\ell}>20~\mbox{GeV},&&~|\eta_{b,\ell}|<2.5,~p_T^j>25 \mbox{GeV},~|\eta_j|>2.5,\nonumber\\ 
 &&\Delta R_{jj,j\ell}>0.4.
\end{eqnarray}
Note that the cut $|\eta_j|>2.5$ corresponds to a very forward light jet which is a characteristic signature of $thj$ process and is instrumental in suppressing the background efficiently. 

\section{Top polarization and angular distributions of the charged
lepton}\label{pol}
\begin{figure}[h!]
\begin{center}
\includegraphics[width=0.4\textwidth]{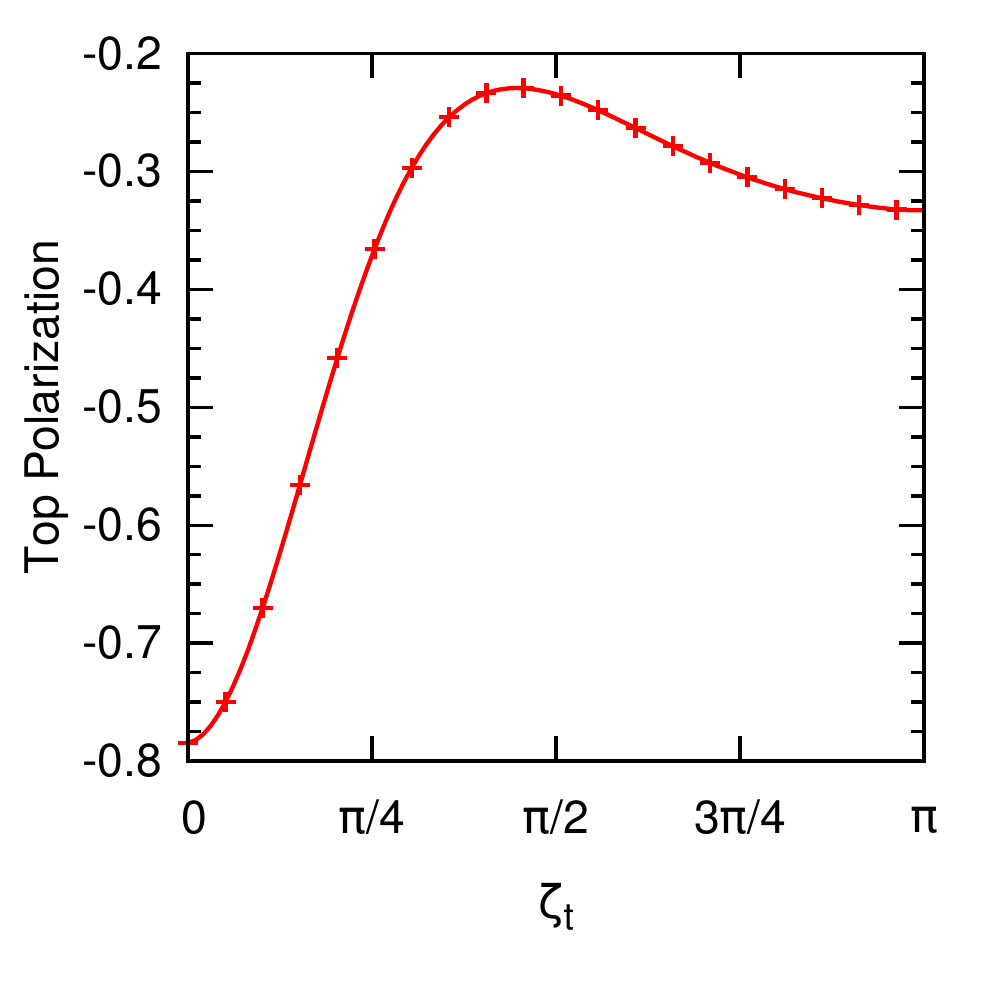}
\caption{ Top polarization in $pp\to thj$  at LHC14 as a function of the CP phase $\zeta_t$ of the $tth$ coupling. } 
\label{pol-top}
\end{center}
\end{figure}

The degree of longitudinal polarization $P_t$ of the top quark is given by
\begin{equation}
P_t=\frac{\sigma(+)-\sigma(-)}{\sigma(+)+\sigma(-)}.
\label{eta3def}
\end{equation}
where $\sigma(+)$ and $\sigma(-)$ denote the cross sections for positive- and negative-helicity top quarks, respectively. The sum of $\sigma(+)$ and $\sigma(-)$ gives the total cross section for the process. We have obtained the polarized cross sections $\sigma(+)$ and $\sigma(-)$ using the helicity amplitudes in {\tt MadGraph}. In Fig.~\ref{pol-top} we display the polarization of the top in $pp\to thj$ at the LHC as a function of the CP phase $\zeta_t$. One can easily see that the polarization is quite sensitive to low values of $\zeta_t$, i.e., $\zeta_t<\pi/2$. This is because of the pseudoscalar coupling which flips the helicity of the top quark in the production amplitude. As the pseudoscalar component in the Higgs admixture is increased with increase in $\zeta_t$, it is expected that the polarization of the top quark would also be affected accordingly. Had there been only one diagram where the Higgs is emitted from the top, the polarization curve would be symmetric around $\zeta_t = \pi/2$ because we retrieve the same CP admixture as in the range (0, $\pi/2$). However, the presence of the second diagram and its interference with the first one results in the flattening of the polarization curve beyond $\zeta_t > \pi/2$.

In the rest frame of the top quark, the angular distribution of a decay product $f$ for a top-quark ensemble has the form 
 \begin{equation}
 \frac{1}{\Gamma_f}\frac{\textrm{d}\Gamma_f}{\textrm{d} \cos \theta _f}
 =\frac{1}{2}(1+\kappa _f P_t \cos \theta _f).
 \label{topdecaywidth}  
 \end{equation}
Here $\theta_f$ is the angle between $f$ and the top spin vector in the top rest frame and $P_t$ (defined in Eq.~(\ref{eta3def})) is the degree of polarization of the top-quark ensemble. $\Gamma_f$ is the partial decay width. The standard way to meaure top polarization is through the angular distribution of its decay products in the rest frame of the top quark, in particular, through the charged lepton and down-type quark distribution whose spin-anlaysing powers $\kappa_\ell=\kappa_d\sim 1$ are maximum while $\kappa_{\nu_\ell}=\kappa_{u}=-0.30$ and $\kappa_{b}=-\kappa_{W^+}=-0.39$.\footnote{All $\kappa$ values are evaluated at tree level \cite{Bernreuther:2008ju}.} A larger $\kappa_f$ makes $f$ a more sensitive probe of the top spin. Thus the $\ell^+$ or $d$ have the largest probability of being emitted in the direction of the top spin and the least probability in the direction opposite to the spin. Since at the LHC, the lepton energy and momentum can be measured with high precision, we focus on leptonic decays of the top. 

As mentioned earlier, the standard way to determine top polarization is to study the charged-lepton polar distribution in the top-quark rest frame, Eq.~\ref{topdecaywidth}. However, this needs a full reconstruction of the top momentum which is a difficult task at the LHC. Utilizing the $W$-boson on-shell condition: $(p_{\ell^\pm}+p_\nu)^2=M_W^2$, one can obtain a quadratic equation in the longitudinal component of neutrino momentum $p_{\nu L}$. Solving this equation, we determine the missing information about $p_{\nu L}$ which brings in a two-fold ambiguity and may thus lead to a considerable loss in the number of events. This becomes even more significant for the case of rare processes like the one under consideration.  

\begin{figure}[h!]
\begin{center}
\includegraphics[width=0.4\textwidth]{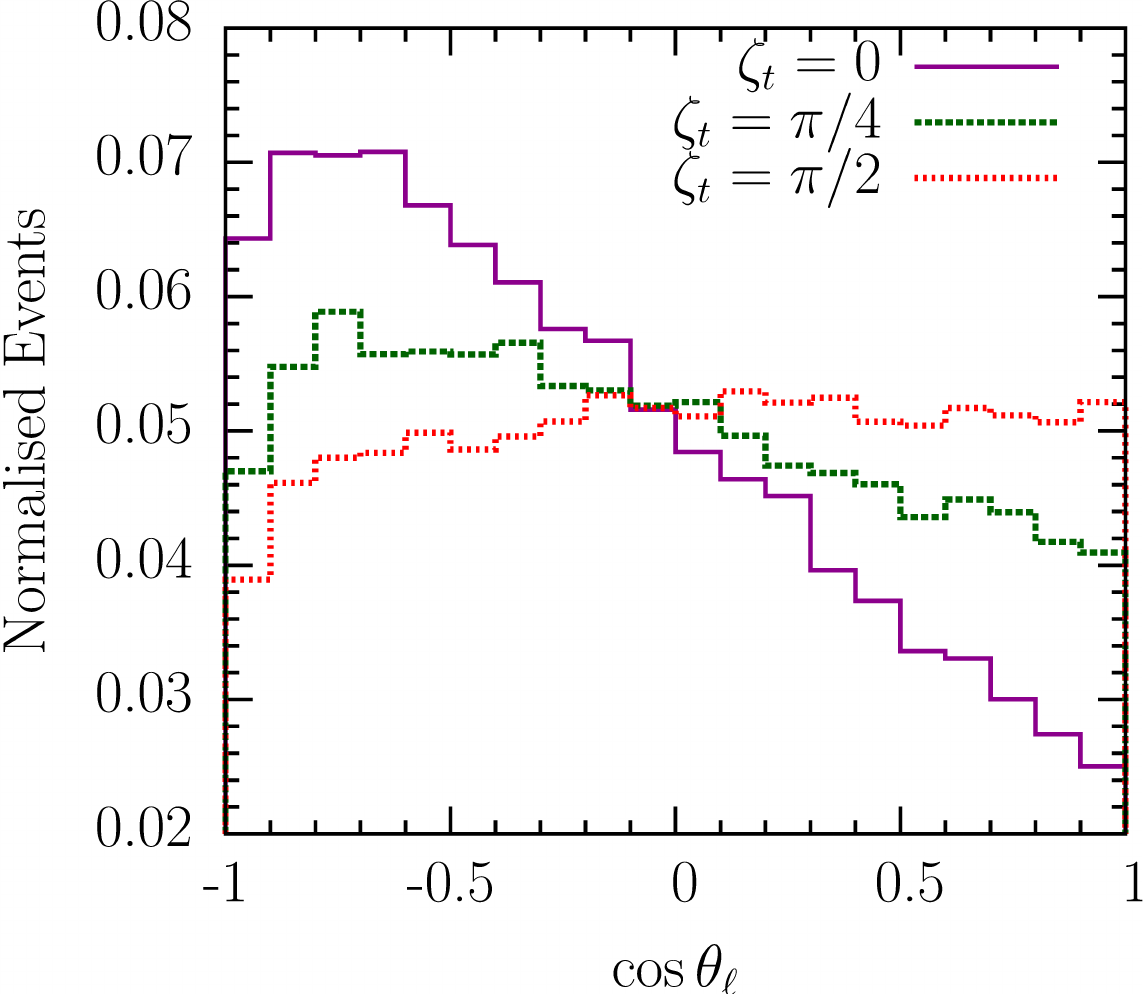}
\includegraphics[width=0.4\textwidth]{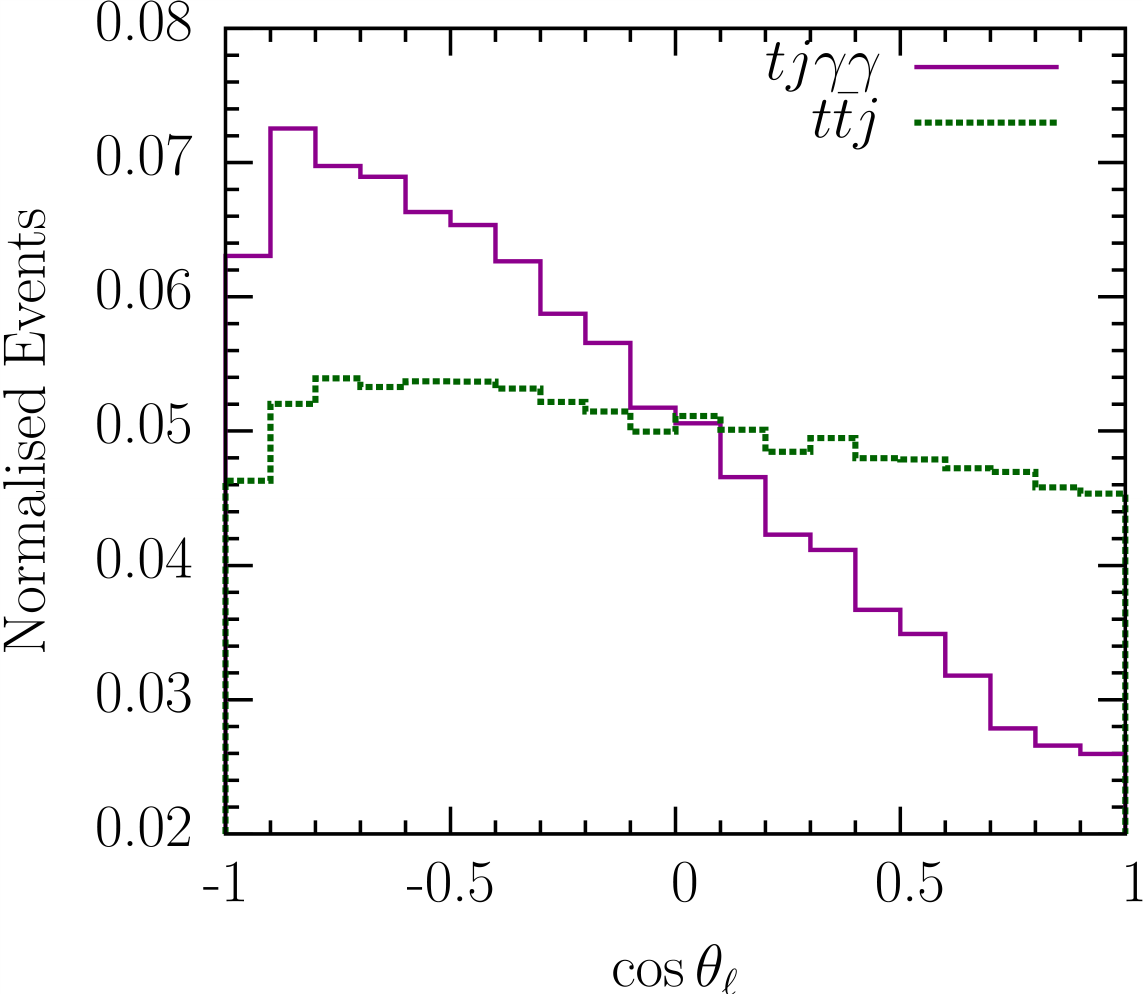}
\caption{ The normalized polar distribution, $\cos\theta_\ell$, of the charged lepton in the top-quark rest frame for $pp\to thj$ (upper panel) process for different values of CP phase $\zeta_t$ of the $t\bar th$ couplings and backgrounds (lower panel) at LHC14.} 
\label{dist-ctl}
\end{center}
\end{figure}

We show in Fig.~\ref{dist-ctl} (upper panel) the normalized distribution in $\cos\theta_{\ell}$, where $\theta_{\ell}$ is the polar angle of the lepton measured with respect to the top-quark spin direction in the rest frame of the top quark, for $pp\to thj$ at LHC14 for two values of anomalous CP phases in $t\bar th$ couplings. Also shown is the distribution for the case of the SM. It can be seen that the top polarization, as measured by the slope of the $\cos\theta_{\ell}$ distribution, is sensitive to the phase $\zeta_t$ of the top Yukawa coupling. We also show, in Fig.~\ref{dist-ctl} (lower panel), the $\cos\theta_\ell$ distribution for processes $t\bar t j$ and $tj\gamma\gamma$ which are the main backgrounds for $pp\to thj, (h\to b\bar b)$ and $pp\to thj, (h\to \gamma\gamma)$ signals respectively. The $t\bar tj$ production is a strong process conserving parity. Hence it leads to vanishing polarization which can be visualized through the flat distribution while $tj\gamma\gamma$ production is mostly electroweak and gives rise to highly polarized tops as evident in the Fig.~\ref{dist-ctl} (lower panel). In order to reconstruct top rest frame, as mentioned earlier, we determine the neutrino longitudinal momentum $p_{\nu L}$ by imposing the invariant mass constraint $M_{l\nu}^2 = M_{W^\pm}^2$ :
\begin{equation}
 p_{\nu L}=\frac{1}{2p_{\ell T}^2}\left(A_W p_{\ell L} \pm E_\ell \sqrt{A_W^2\pm 4 p_{\ell T}^2 \MET^2}\right),
\end{equation}
where $A_W=M_{W^\pm}^2+2\vec{p}_T\cdot { \vec{\rm E}{\!\!\!/}_T}$. If two solutions for $p_{\nu L}$ are found, the one which gives $M_{l\nu}$ closer to the $W^\pm$ mass is adopted. Also, we reject the events with complex solutions.
 
In order to avoid difficulties associated with the reconstruction of the top rest frame, we consider an observable that can be measured directly in the lab frame, {\it viz.}, the azimuthal distribution of the charged lepton arising from top decay. To define the azimuthal angle $\phi_\ell$, we choose the proton beam direction as the $z$ direction, and the production plane of the top quark as the $xz$ plane. The measurement of $\phi_\ell$ does not need full reconstruction of the top momentum, but only the transverse momentum of top quark.

The angular distribution, analogous to Eq.~\ref{topdecaywidth}, in the lab frame in terms of angle $\theta_{t\ell}$ between the top and lepton directions can be written as \cite{Godbole:2010kr}
\beq\label{labdist}  \displaystyle
\frac{1}{\Gamma_{\ell}}\frac{d\Gamma_{\ell}}{d\cos\theta_{t\ell}} = \displaystyle \frac{1}{2}
(1-\beta^2)(1 - P_t \beta)\frac{1 + P_t^{\rm eff} \cos\theta_{t\ell}}{(1- \beta
\cos\theta_{t\ell})^3},
\eeq
where $\beta = \sqrt{1 - m_t^2/E_t^2}$,  
\beq\label{costhetatl}
\cos\theta_{t\ell} = \cos\theta_t \cos\theta_{\ell} + \sin\theta_t \sin\theta_{\ell}
\cos\phi_{\ell};
\eeq
and 
\beq
P_t^{\rm eff}=\frac{P_t - \beta}{1 - P_t \beta}
\eeq

\begin{figure}[h!]
\begin{center}
\includegraphics[width=0.4\textwidth]{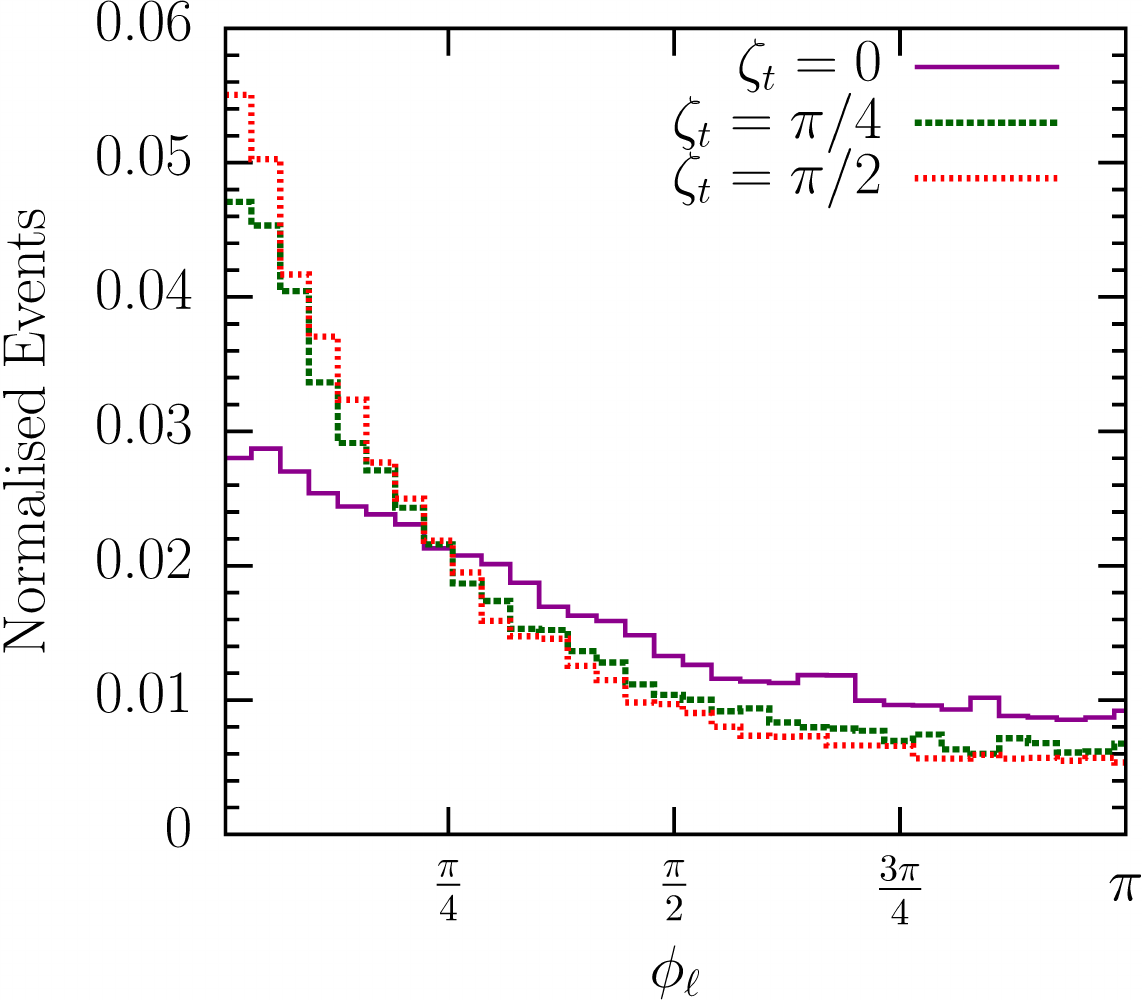}
\includegraphics[width=0.4\textwidth]{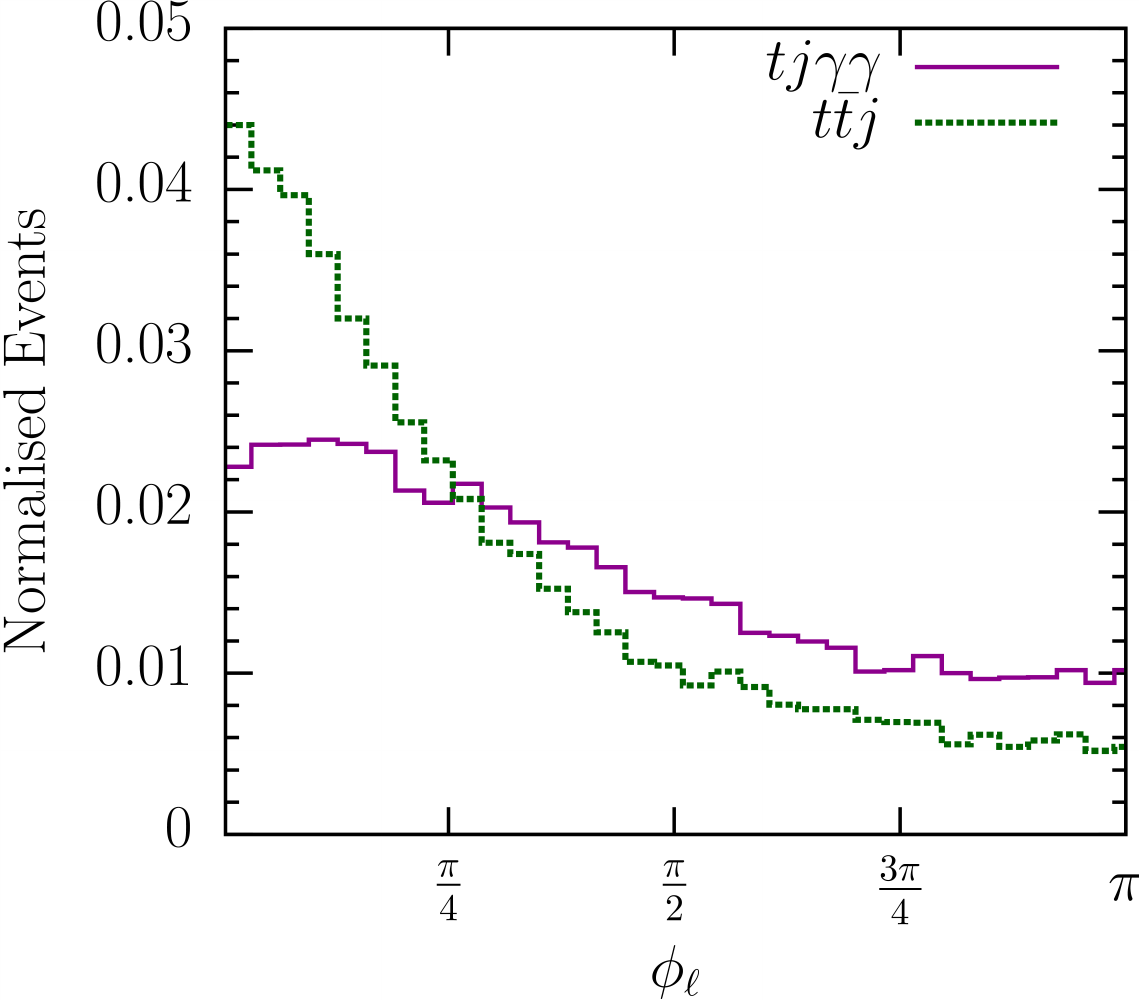}
\caption{ The normalized distribution in the azimuthal angle $\phi_\ell$ of the charged lepton in $thj$ production (upper panel) for different values of the CP phase $\zeta_t$ in the $t\bar th$ coupling and in the background processes (lower panel) at LHC14.} 
\label{dist-phi}
\end{center}
\end{figure}

Thus, the azimuthal distribution not only depends on polarization of top but also on a kinematic effect. According to Eq.~\ref{topdecaywidth}, the decay lepton is emitted preferentially along the top spin direction in the top rest frame, with $\kappa_f=1$. The corresponding distributions in the lab frame are given by Eq.~\ref{labdist}. The rest-frame forward (backward) peak corresponds to a peak for $\cos\theta_{t\ell}=\pm 1$, as seen from the factor $(1 + P_t^{\rm eff} \cos\theta_{t\ell})$ in the numerator of Eq.~\ref{labdist}. This is the effect of polarization. The kinematic effect is seen in the factor $(1 - \beta_t \cos \theta_{t\ell})^3$ in the denominator of Eq. \ref{labdist} , which again gives rise to peaking for large $\cos\theta_{t\ell}$. Eq.~\ref{costhetatl} therefore implies peaking for small $\phi_{\ell}$. This is borne out by the numerical results. 

We show in Fig.~\ref{dist-phi} the normalized azimuthal distribution of the charged lepton in $thj$ production at LHC14 for a few values of $\zeta_t$, $\zeta_t=0$ corresponding to the SM. As expected and as can be seen from the figure, the distribution is sensitive to $\zeta_t$. We also show in Fig.~\ref{dist-phi} (bottom), the $\phi_\ell$ distribution for processes $t\bar t j$ and $tj\gamma\gamma$ which are the main backgrounds for $pp\to thj, (h\to b\bar b)$ and $pp\to thj, (h\to \gamma\gamma)$ signals respectively. The $\phi_\ell$ distribution in Fig.~\ref{dist-phi} is symmetric under the interchange of $\phi_\ell$ with $2\pi-\phi_\ell$. This is because of the fact that the LHC is a symmetric collider and there is no way to define a unique positive $z$ axis. In Fig.~\ref{dist-phi} we have shown the distribution only up to $\pi$.

The lab-frame charged-lepton azimuthal distribution as a probe of top-quark polarization was first proposed in Ref.~\cite{Godbole:2006tq}. Subsequently, it has been studied extensively in the context of various new-physics scenarios in processes involving top pair production \cite{Godbole:2010kr,Godbole:2009dp,Biswal:2012dr} and (associated) single-top production \cite{Huitu:2010ad,Godbole:2011vw,Rindani:2011pk,Rindani:2011gt,Rindani:2013mqa,Rindani:2015vya,Rindani:2015dom,Rindani:2015bhu} at the LHC. 

\section{Asymmetries}\label{sec-asym}

As seen in the previous section, one can use polar and azimuthal angular distributions of the charged lepton to discriminate amongst possible values of the top Yukawa phase $\zeta_t$. However, making a fit to the distributions requires a reasonably large data sample. It is, thus, preferable to compare the data to a single number defined in terms of an integral over the distribution. For this purpose, we define an asymmetry in each of the previous cases, and evaluate it as a function of $\zeta_t$.

We define a polar asymmetry, which is also the forward-backward asymmetry of the charged lepton in rest frame of top quark, by
\begin{equation}
 \mathcal A_{\ell}^{\rm FB}=\frac{\sigma(\cos \theta_{\ell} >0)-\sigma(\cos\theta_{\ell}<0)}{\sigma(\cos \theta_{\ell}>0)+\sigma(\cos\theta_{\ell}<0)},
\label{polasy}
\end{equation}
where, as before $\theta_{\ell}$ is the polar angle of the charged lepton relative to the top spin direction in the top rest frame. 

\begin{figure}[h!]
\begin{center}
\includegraphics[width=0.4\textwidth]{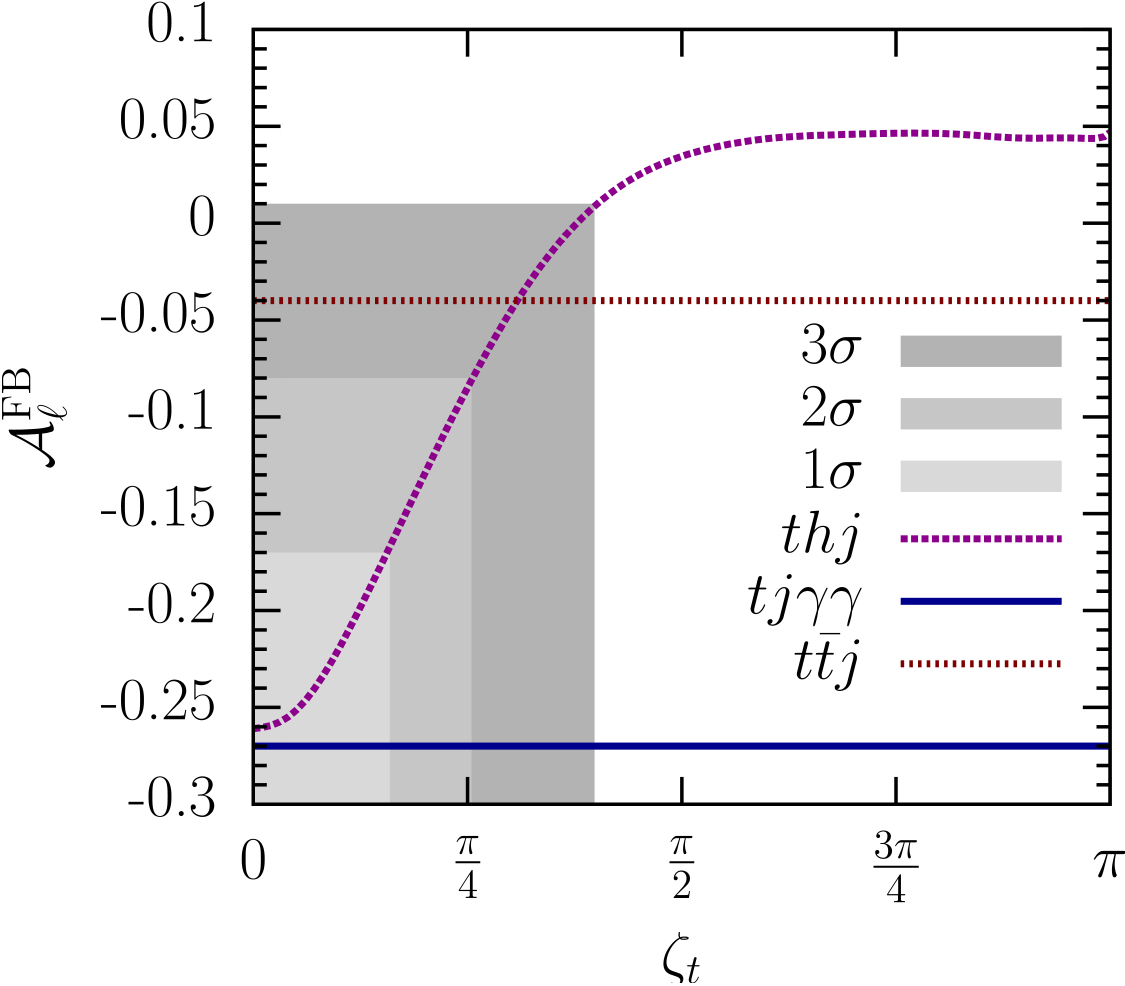}
\includegraphics[width=0.4\textwidth]{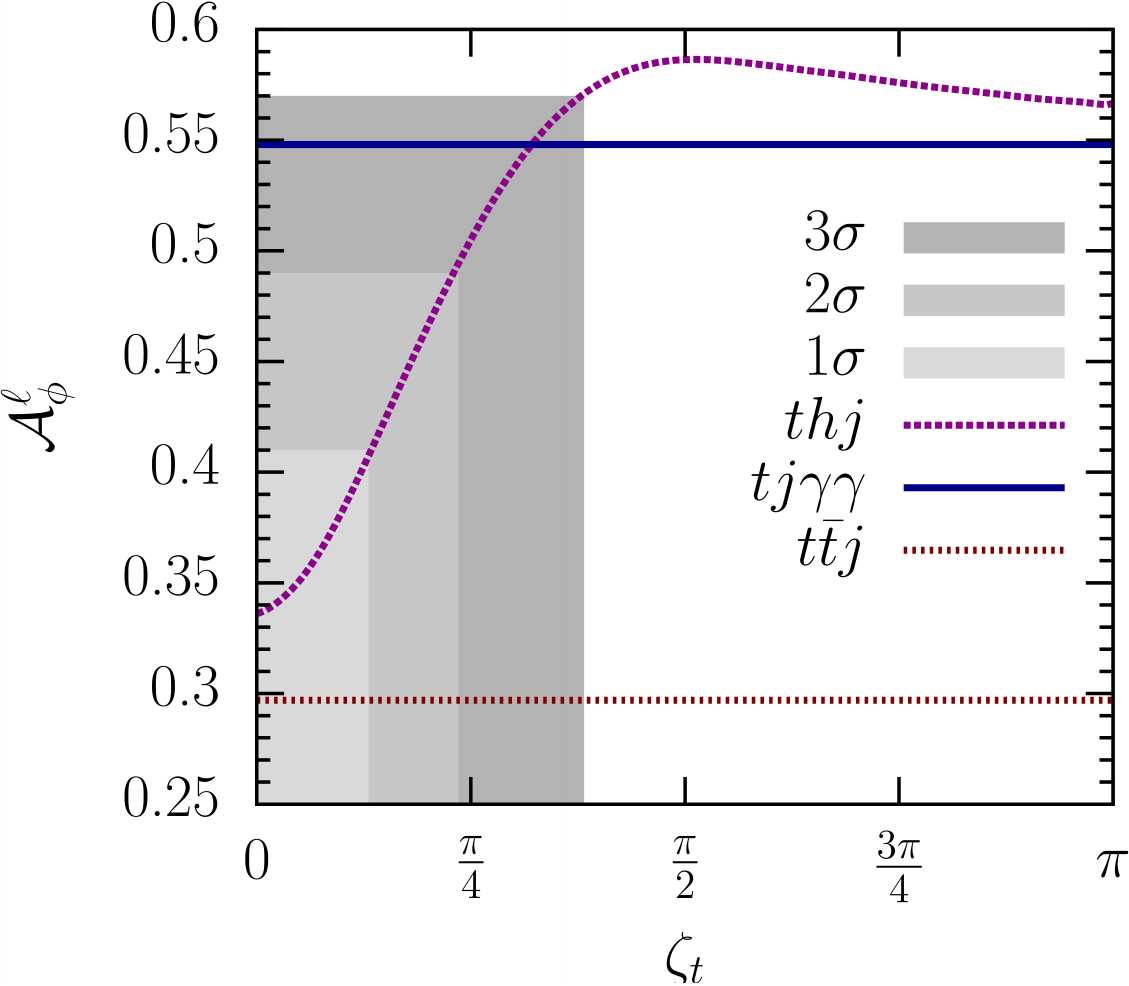}
\caption{ Charged-lepton polar asymmetry ($A_{\ell}^{\rm FB}$) (upper panel) and azimuthal asymmetry ($A_{\phi}^\ell$) (lower panel) in $pp\to thj$  at LHC14 as a function of the CP phase $\zeta_t$ of the $tth$ coupling. Also shown are the values of the asymmetries for the background processes $t\bar tj$ and $tj\gamma\gamma$. The different shades of gray regions denote the 1$\sigma$, 2$\sigma$ and 3$\sigma$ of statistical uncertainty in the mesurement of the asymmetries in the SM. } 
\label{asym}
\end{center}
\end{figure}

In the production plane of the top-quark, we deinfe an azimuthal asymmetry, which is in fact the ``left-right asymmetry'' of the charged lepton at the LHC defined with respect to the beam direction, with the right hemisphere identifed as that in which the top momentum lies, and the left one being the opposite one.  In the Fig.~\ref{dist-phi}, it can be easily seen that the $\phi_\ell$ distribution is highly asymmetric in the two different regions, {\it viz.} left ($\cos\phi_\ell<0$) and right ($\cos\phi_\ell>0$), of the detector. We define the lab frame left-right asymmetry of charged lepton, as follows:
\begin{equation}
 \mathcal A_\phi^\ell=\frac{\sigma(\cos \phi_\ell >0)-\sigma(\cos\phi_\ell<0)}{\sigma(\cos \phi_\ell >0)+\sigma(\cos \phi_\ell<0)},
\label{aziasy}
\end{equation}
where the denominator is the total cross section. 

We also study the sensitivities of these asymmetries as a probe of the CP violating phase at the LHC14 with the full integrated luminosity, {\it viz.}, 3000 fb$^{-1}$. For this, we estimate the statistical uncertainty in the measurement of an asymmetry using the formula 
\begin{equation}\label{sd}
 \Delta \mathcal A=\frac{\sqrt{1-\mathcal A_{\rm SM}^2}}{\sqrt{\sigma_{\rm SM} \mathcal L}},
\end{equation}
where $\mathcal L$, $\mathcal A_{\rm SM}$ and $\sigma_{\rm SM}$ are the integrated luminosity, the value of an asymmetry and the total cross section in the SM respectively.

In Fig.~\ref{asym}, we present the leptonic asymmetries $\mathcal A_{\ell}^{\rm FB}$ and $\mathcal A_\phi^\ell$ as functions of CP phase $\zeta_t$ at LHC14. We can see from the figure that the asymmetry $\mathcal A_\phi^\ell$ reconstructs fairly accurately the behaviour of the top polarization. The top rest-frame polar asymmetry $\mathcal A_{\ell}^{\rm FB}$ also follows the same behaviour, though to a lesser extent. The advantage of $\mathcal A_\phi^\ell$, in addition to having a shape closer to that of the actual polarization, is that it can be measured in the lab frame. Thus we expect better sensitivity to $\zeta_t$ from $\mathcal A_\phi^\ell$ than $\mathcal A_{\ell}^{\rm FB}$. In the Fig.~\ref{asym}, we also show the regions which can be probed with 3000 fb$^{-1}$ of integrated luminosity at 1$\sigma$, 2$\sigma$ and 3$\sigma$ of significance at the 14 TeV LHC. In particular, with a total luminosity of about 3 ab$^{-1}$ likely be available at the end of the HL-LHC run, $A_\phi^\ell$ could be used to determine $\zeta_t$ to within $\pi/8,~\pi/4$ and $3\pi/8$ at 1$\sigma$, 2$\sigma$ and $3\sigma$ confidence level (CL) respectively. 

\section{Conclusions}\label{conclusions}

Post the Higgs discovery, the need of the hour is to determine the CP properties of the Higgs boson unambiguously. The fact that a pseudoscalar does not couple to the EW gauge bosons at tree level spurs the idea of studying the CP properties of the Higgs in fermionic Yukawa couplings as they are more democratic to CP even and odd scalars. Moreover the current measurement of the CP phase in the top Yukawa couplings relies on $h\gamma\gamma$ and $hgg$ couplings which are deduced from a loop-level calculation,  and thus allow contamination from various new physics effects. This compels us to look for direct determination of such couplings at the LHC. The processes which have the putative couplings have very small cross sections and thus would require a high energy and high luminosity run of the LHC to be completed.

In this letter, we have studied the prospects of measuring the CP phase in the top-Higgs coupling in the associated $thj$ production at the LHC. In this context, we utilize a simpler lab-frame asymmetry $A_\phi^\ell$ of the charged lepton from top decay, which is also the left-right asymmetry of the charged lepton, at the LHC. We find that the left-right asymmetry is quite sensitive to the CP violating phase and can probe it up to $\pi/6$ with 3 ab$^{-1}$ of the integrated luminosity at the LHC. We also study the angular distribution of charged-lepton in the top rest frame. The rest-frame forward-backward asymmetry $A_\ell^{\rm FB}$ gives a measure of top-quark polarization in production. However it requires a full reconstruction of top momentum which brings in large systematic uncertainties. Thus the sensitivity of $A_\ell^{\rm FB}$ lesser than $\mathcal A^\ell_\phi$ which only requires the reconstruction of transverse momentum of top quark.

The asymmetries and their sensitivities have been estimated at the parton level though we have employed all the relevant cuts to suppress the signal-to-background ratio. However, including the detector effects may lead to reduction in the sensitivities of these asymmetries. It is thus needed to do perform a full detector level simulation to estimate the realistic efficiencies of these observables. We have left this as a future work.

\section*{Acknowledgments}
SDR acknowledges support from the Department of Science and Technology, India, under the J.C. Bose National Fellowship programme, Grant No. SR/SB/JCB-42/2009. The work of P.S. was supported by the University of Adelaide and the Australian Research Council through the ARC Center of Excellence in Particle Physics (CoEPP) at the Terascale (CE110001004).

\bibliography{bibliography_tth}
\end{document}